\documentclass[aps,superscriptaddress,preprint]{revtex4-1}
\usepackage{amsmath}
\usepackage{amssymb}
\usepackage{graphicx}
\usepackage[colorlinks,citecolor=blue, linkcolor=blue,hyperindex,CJKbookmarks,dvipdfm]{hyperref}

\begin{document}

\title{Nonreciprocal transition between two indirectly coupled energy levels}
\author{Xun-Wei Xu}
\email{davidxu0816@163.com}
\affiliation{Key Laboratory of Low-Dimensional Quantum Structures and Quantum Control of Ministry of Education, Key Laboratory for Matter Microstructure and Function of Hunan Province, Department of Physics and Synergetic Innovation Center for Quantum Effects and Applications, Hunan Normal University, Changsha 410081, China}
\affiliation{Department of Applied Physics, East China Jiaotong University, Nanchang,
330013, China}
\author{Hai-Quan Shi}
\affiliation{Department of Applied Physics, East China Jiaotong University, Nanchang,
330013, China}
\author{Ai-Xi Chen}
\email{aixichen@zstu.edu.cn}
\affiliation{Department of Physics, Zhejiang Sci-Tech University, Hangzhou
310018, China}
\affiliation{Department of Applied Physics, East China Jiaotong University, Nanchang,
330013, China}

\date{\today }


\begin{abstract}
We propose a theoretical scheme to realize nonreciprocal transition between two energy levels that can not coupled directly. Suppose they are coupled indirectly by two auxiliary levels with a cyclic four-level configuration, and the four transitions in the cyclic configuration are controlled by external fields. The indirectly transition become nonreciprocal when the time reversal symmetry of the system is broken by the synthetic magnetic flux, i.e., the total phase of the external driving fields through the cyclic four-level configuration.
The nonreciprocal transition can be identified by the elimination of a spectral line in the spontaneous emission spectrum. Our work introduces a feasible way to observe nonreciprocal transition in a wide range of multi-level systems, including natural atoms or ions with parity symmetry.
\end{abstract}

\maketitle

\section{Introduction}

Time reversal symmetry is the hypothesis that certain physical quantities are unchanged under time reversal transformation, which is related to reversibility of the system, such as the principle of detailed balancing in kinetic systems~\cite{Pathria}.
With the principle of detailed balance as a background, A. Einstein proposed his quantum theory of radiation~\cite{Einstein1917} in 1916, which is now considered as the theoretical foundation of the
laser, and one of the important corollaries is the absorption coefficient should be equal to the stimulated emission coefficient between two nondegenerate energy levels.
However, sometimes we need to break the time reversal symmetry of atomic systems to yield fantastic phenomena, e.g., cyclic population transfer~\cite{KralPRL01}, controllable electromagnetically induced transparency~\cite{HLiPRA09}, gain without
inversion~\cite{WZJiaPRA10}.

One important approach to break the time reversal symmetry of atomic systems is based on the cyclic three-level transitions in the chiral molecules~\cite{KralPRL01,KralPRL03,YLiPRL07}, and in the superconducting qubit circuit with three Josephson junctions~\cite{YXLiuPRL05,MooijSci99}.
The cyclic three-level transitions in multi-level atomic systems have been used to generate many interesting phenomena in single-photon level, including single-photon quantum routing~\cite{LZhouPRL13}, single-photon second-order nonlinear processes~\cite{YXLiuSR14,YJZhaoPRA17,ZHWangPRA15}.
When the three possible transitions in the cyclic three-level configuration are driven by three mutually phase-locked driving fields, the time reversal symmetry of atomic system can be broken by the magnetic flux synthesized from the driving-field phase.
In a recent experiment, time-reversal symmetry breaking and cyclic population transfer were revealed in a single nitrogen-vacancy spin system by controlling the global phase of the driving fields~\cite{BarfussNPy18}.

On the basis above, a multi-level atomic system with cyclic three-level configuration was proposed to realize significant difference between the stimulated emission and absorption coefficients of two nondegenerate energy levels~\cite{XWXuArx19}, which was referred to as nonreciprocal transition.
Different from the closed-contour spin dynamics~\cite{BarfussNPy18}, besides synthetic magnetism, reservoir engineering was also employed to eliminate one of the transitions in opposite directions~\cite{XWXuArx19}.
Nonreciprocal transition in the multi-level atomic system provides us a new physical mechanism for designing nonreciprocal photon devices~\cite{JZhangPRB15,YJiangPRAPP18} at single-photon level with single atoms~\cite{XWXuArx19}.
Moreover, it was shown that the nonreciprocal transition can lead to the elimination of a spectral line in the spontaneous emission spectrum~\cite{XWXuArx20}, which has a potential application for the determination of enantiomeric excess of chiral molecules~\cite{XWXuArx20}.

It is well known that due to the parity symmetry of the potential energy for usual natural atomic systems, described by $SO(3)$ or $SO(4)$, the parities of atomic eigenstates are well defined, and one-photon transitions between two energy levels require that the two corresponding eigenstates have opposite parities, which is referred to as the selection rule for the electric-dipole transitions. In a three-level atom system, at least two of the atomic eigenstates have the same parity and the one-photon dipole transition between them is forbidden. Thus, a cyclic three-level atom, i.e., the population cyclically transferred between three energy levels, cannot be realized with three one-photon transitions, except that the parity symmetry of the atomic system is broken. We note that the parity symmetry of a natural atom can be broken by applying a strong magnetic field~\cite{AnsariPRA90,BlockleyPRA91}. However, this technique is difficult to implement and isn't commonly used.

In this paper, we propose a theoretical scheme to realize nonreciprocal transition between two energy levels that cannot coupled directly, which is significantly different from the models in Refs.~\cite{XWXuArx19,XWXuArx20}.
Most important, this provides a way to realize nonreciprocal transition without breaking the parity symmetry of the system, which open up a feasible way to observe nonreciprocal transition in wider systems, such as natural atoms or ions.
Moreover, as two energy levels for nonreciprocal transition are not coupled directly, so there is not any strict restriction on the energy difference between them, and nonreciprocal transition can be realized between two degenerate energy levels, which goes beyond the limit of nondegenerate energy levels in Refs.~\cite{XWXuArx19,XWXuArx20}.
Our model can be used to design atom- or ion-mediated nonreciprocal devices with well-established technologies in the fields of cold atom~\cite{RitschRMP13} and ion trap~\cite{BrownnuttRMP15,TomzaRMP19}. Such devices can find applications for quantum control of light in chiral quantum technologies~\cite{LodahlNarure17} or topological photonics~\cite{OzawaRMP19}.

The remainder of this paper is organized as follows. In Sec.~\ref{HD}, a cyclic four-level configuration for two energy levels coupled indirectly by two auxiliary levels is introduced and the dynamical equations are given under the Weisskopf-Wigner approximation. The time evolution of the
populations and the transition probabilities between the two energy levels are investigated, and nonreciprocal transitions are shown in
Sec.~\ref{TP}. Moreover, the spontaneous emission spectra of the systems with nonreciprocal transitions are discussed in Sec.~\ref{SE}, which provide us a convenient way to measure nonreciprocal transitions in experiments.
Finally, the conclusions is given in Sec.~\ref{CN}.

\section{Hamiltonian and dynamical equations} \label{HD}

\begin{figure}[tbp]
\centering
\includegraphics[bb=80 314 527 689, width=7.5 cm, clip]{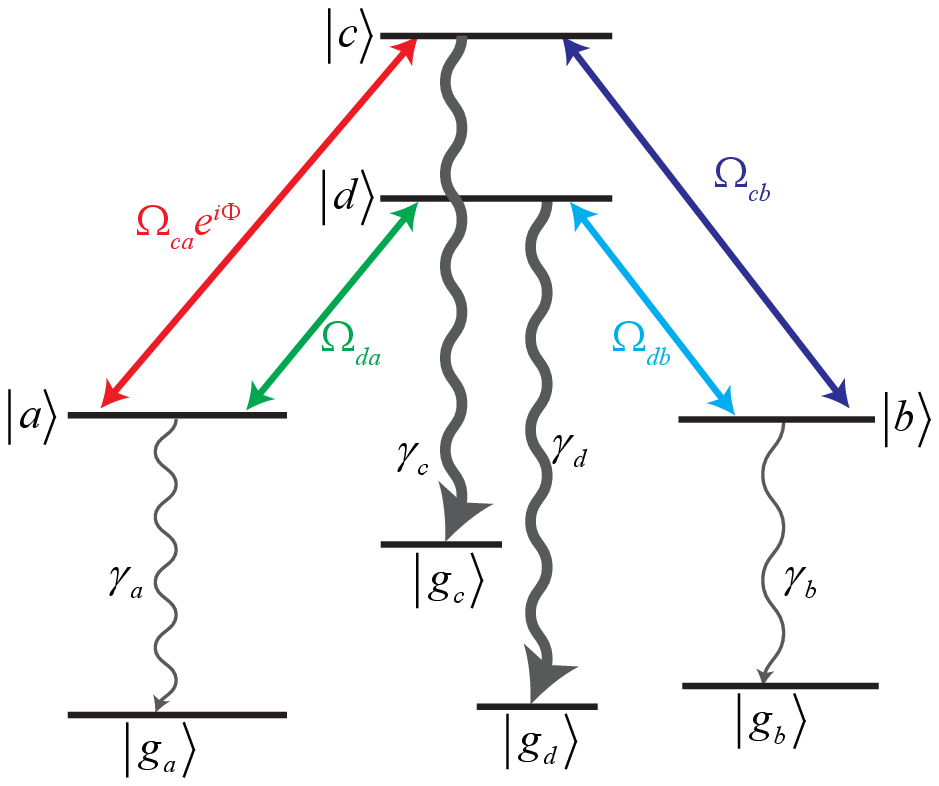}
\caption{(Color online) Level diagram of an atom or ion with cyclic transitions for
the four upper levels ($|a\rangle$, $|b\rangle$, $|c\rangle$, and $|d\rangle$), and they
are coupled by the same vacuum modes to different lower levels ($|g_a\rangle$, $|g_b\rangle$, $|g_c\rangle$, and
$|g_d\rangle$).
}
\label{fig1}
\end{figure}

We propose to realize nonreciprocal transition between two energy levels $|a\rangle$ and $|b\rangle$,
under the assumption that they can not coupled directly.
Suppose they are coupled indirectly through two auxiliary levels ($|c\rangle$ and $|d\rangle$) as a cyclic four-level configuration,
and the four transitions are controlled by external fields with frequencies ($\nu_{ca}$, $\nu_{cb}$, $\nu_{db}$, and $\nu_{da}$), Rabi
frequencies ($\Omega_{ca}$, $\Omega_{cb}$, $\Omega_{db}$, and $\Omega_{da}$) and phases ($%
\phi_{ca}$, $\phi_{cb}$, $\phi_{db}$, and $\phi_{da}$), as shown in Fig.~%
\ref{fig1}. For different systems, the four (upper) levels may be coupled to the same lower level or to different lower levels respectively. Previous studies have shown that the spontaneous emission from multiple upper levels to the common lower level may result in spontaneous emission cancellation and spectral line elimination~\cite%
{SYZhuPRA95,SYZhuPRL96}, which is not the focus of this paper. In order to eliminate this effect, the spontaneous emission spectrum for the system will be derived for the case that the four upper levels
are coupled to four different lower levels ($|g_a\rangle$, $|g_b\rangle$, $|g_c\rangle$,
and $|g_d\rangle$) respectively with the same vacuum modes.
The Hamiltonian is given by ($\hbar =1$)%
\begin{eqnarray}
H &=&\Omega _{ca}e^{i\phi _{ca}}e^{-i\Delta _{ca}t}\left\vert a\right\rangle
\left\langle c\right\vert +\Omega _{cb}e^{i\phi _{cb}}e^{i\Delta
_{cb}t}\left\vert c\right\rangle \left\langle b\right\vert \nonumber\\
&&+\Omega
_{da}e^{i\phi _{da}}e^{i\Delta _{da}t}\left\vert d\right\rangle \left\langle
a\right\vert +\Omega _{db}e^{i\phi _{db}}e^{-i\Delta _{db}t}\left\vert
b\right\rangle \left\langle d\right\vert  \nonumber\\
&&+\sum_{k}\left[ g_{k}^{a}e^{i\left( \omega _{ag}-\omega _{k}\right)
t}v_{k}\left\vert a\right\rangle \left\langle g_{a}\right\vert
+g_{k}^{b}e^{i\left( \omega _{bg}-\omega _{k}\right) t}v_{k}\left\vert
b\right\rangle \left\langle g_{b}\right\vert\right] \nonumber\\
&&+\sum_{k}\left[ g_{k}^{c}e^{i\left( \omega
_{cg}-\omega _{k}\right) t}v_{k}\left\vert c\right\rangle \left\langle
g_{c}\right\vert +g_{k}^{d}e^{i\left( \omega _{dg}-\omega _{k}\right)
t}v_{k}\left\vert d\right\rangle \left\langle g_{d}\right\vert \right]  \nonumber\\
&&+\mathrm{H.c.,}
\end{eqnarray}
where $v_{k}$ and $v_{k}^{\dagger }$ are annihilation and creation operators for photons in the $k$th vacuum mode with frequency $%
\omega _{k}$ ($k$ denotes both the momentum and polarization of the
vacuum modes); $\omega _{ij}$ and $\omega _{ig}$ are the frequency differences between levels $\left\vert i\right\rangle $ and $%
\left\vert j\right\rangle $, and between $\left\vert i\right\rangle $ and $\left\vert g_i\right\rangle $ respectively, with $\left( i,j=a,b,c,d\right)$; $\Delta _{ij}\equiv\omega _{ij}-\nu _{ij}$ is the
detuning between the atomic transition $\left\vert i\right\rangle \leftrightarrow \left\vert j\right\rangle$ and the driving field, and $g_{k}^{i}$ is the coupling strength between the atomic transition $|i \rangle \leftrightarrow |g_i\rangle $ and the $k$th vacuum mode.
With the replacement $\left\vert b\right\rangle \rightarrow e^{i\phi _{cb}}\left\vert
b\right\rangle $, $\left\vert d\right\rangle \rightarrow e^{i\left( \phi
_{db}+\phi _{cb}\right) }\left\vert d\right\rangle $, $\left\vert
a\right\rangle \rightarrow e^{i\left( \phi _{da}+\phi _{db}+\phi
_{cb}\right) }\left\vert a\right\rangle $, $g_{k}^{b}\rightarrow
g_{k}^{b}e^{-i\phi _{cb}}$, $g_{k}^{d}\rightarrow g_{k}^{d}e^{-i\left( \phi
_{db}+\phi _{cb}\right) }$, and $g_{k}^{a}\rightarrow g_{k}^{a}e^{-i\left(
\phi _{da}+\phi _{db}+\phi _{cb}\right) }$, the Hamiltonian is rewritten as
\begin{eqnarray}\label{Eq1}
H &=&\Omega _{ca}e^{i\Phi }e^{-i\Delta _{ca}t}\left\vert a\right\rangle
\left\langle c\right\vert +\Omega _{cb}e^{i\Delta _{cb}t}\left\vert
c\right\rangle \left\langle b\right\vert
\nonumber \\
&& +\Omega _{da}e^{i\Delta
_{da}t}\left\vert d\right\rangle \left\langle a\right\vert +\Omega
_{db}e^{-i\Delta _{db}t}\left\vert b\right\rangle \left\langle d\right\vert
\nonumber \\
&&+\sum_{k}\left[ g_{k}^{a}e^{i\left( \omega _{ag}-\omega _{k}\right)
t}v_{k}\left\vert a\right\rangle \left\langle g_{a}\right\vert
+g_{k}^{b}e^{i\left( \omega _{bg}-\omega _{k}\right) t}v_{k}\left\vert
b\right\rangle \left\langle g_{b}\right\vert\right]
\nonumber \\
&&+\sum_{k}\left[ g_{k}^{c}e^{i\left( \omega
_{cg}-\omega _{k}\right) t}v_{k}\left\vert c\right\rangle \left\langle
g_{c}\right\vert +g_{k}^{d}e^{i\left( \omega _{dg}-\omega _{k}\right)
t}v_{k}\left\vert d\right\rangle \left\langle g_{d}\right\vert \right]
\nonumber \\
&&+\mathrm{H.c.,}
\end{eqnarray}%
where $\Phi \equiv \phi _{ca}+\phi _{db}+\phi _{cb}+\phi _{da}$ is the total phase of the four strong driving fields through the cycle-transition $|a\rangle \rightarrow |d\rangle \rightarrow |b\rangle \rightarrow |c\rangle \rightarrow |a\rangle $, i.e., the synthetic magnetic flux. Moreover, real coupling strength $g_{k}^{i}$ is assumed for the phase of $g_{k}^{i}$ does not matter in the following discussions. For simplicity, we also make the assumption of resonance $\Delta _{c}\equiv \Delta _{ac}=\Delta _{bc}$ and $\Delta _{d}\equiv\Delta _{da}=\Delta _{db}$.

In the following, we will discuss the indirect transition between levels $|a\rangle $ and $|b\rangle $ based on the Schr\"{o}dinger equation under the Weisskopf-Wigner approximation~\cite{WW30,SZ30}.
We assume that the system is prepared in level $|a\rangle $ or $|b\rangle $ initially, i.e., $|\psi(0)\rangle=|a\rangle|0\rangle$ or $|\psi(0)\rangle=|b\rangle|0\rangle$ with $|0\rangle$ denoting the vacuum state. According to the Schr\"{o}dinger equation, $d\left\vert \psi (t) \right\rangle /dt=-iH\left\vert \psi(t)
\right\rangle $, the state vector at time $t$ can be written as%
\begin{equation}
\left\vert \psi \left( t\right) \right\rangle =\left[ A\left( t\right)
\left\vert a\right\rangle +B\left( t\right) \left\vert b\right\rangle
+C\left( t\right) \left\vert c\right\rangle +D\left( t\right) \left\vert
d\right\rangle \right] \left\vert 0\right\rangle
+\sum_{i=a,b,c,d}\sum_{k}G_{k}^{i}\left( t\right) v_{k}^{\dagger } \left\vert
g_{i}\right\rangle \left\vert 0\right\rangle ,
\end{equation}%
with occupation populations $|A(t)|^2$, $|B(t)|^2$, $|C(t)|^2$, $|D(t)|^2$, and $|G_{k}^{i}(t)|^2$ in the corresponding levels. Under the
Weisskopf-Wigner approximation~\cite{WW30,SZ30}, the dynamical behaviors for the coefficients are obtained as
\begin{equation}\label{Eq3}
\frac{d}{dt}A\left( t\right) =-\frac{\gamma _{a}}{2}A\left( t\right)
-i\Omega _{ca}e^{i\Phi }e^{-i\Delta _{c}t}C\left( t\right) -i\Omega
_{da}e^{-i\Delta _{d}t}D\left( t\right),
\end{equation}%
\begin{equation}\label{Eq4}
\frac{d}{dt}B\left( t\right) =-\frac{\gamma _{b}}{2}B\left( t\right)
-i\Omega _{cb}e^{-i\Delta _{c}t}C\left( t\right) -i\Omega _{db}e^{-i\Delta
_{d}t}D\left( t\right),
\end{equation}%
\begin{equation}\label{Eq5}
\frac{d}{dt}C\left( t\right) =-\frac{\gamma _{c}}{2}C\left( t\right)
-i\Omega _{ca}e^{-i\Phi }e^{i\Delta _{c}t}A\left( t\right) -i\Omega
_{cb}e^{i\Delta _{c}t}B\left( t\right),
\end{equation}%
\begin{equation}\label{Eq6}
\frac{d}{dt}D\left( t\right) =-\frac{\gamma _{d}}{2}D\left( t\right)
-i\Omega _{da}e^{i\Delta _{d}t}A\left( t\right) -i\Omega _{db}e^{i\Delta
_{d}t}B\left( t\right),
\end{equation}%
\begin{equation}\label{Eq7}
\frac{d}{dt}G_{k}^{a}\left( t\right) =-ig_{k}^{a}e^{-i\left( \omega
_{ag}-\omega _{k}\right) t}A\left( t\right),
\end{equation}%
\begin{equation}\label{Eq8}
\frac{d}{dt}G_{k}^{b}\left( t\right) =-ig_{k}^{b}e^{-i\left( \omega
_{bg}-\omega _{k}\right) t}B\left( t\right),
\end{equation}%
\begin{equation}\label{Eq9}
\frac{d}{dt}G_{k}^{c}\left( t\right) =-ig_{k}^{c}e^{-i\left( \omega
_{cg}-\omega _{k}\right) t}C\left( t\right),
\end{equation}%
\begin{equation}\label{Eq10}
\frac{d}{dt}G_{k}^{d}\left( t\right) =-ig_{k}^{d}e^{-i\left( \omega
_{dg}-\omega _{k}\right) t}D\left( t\right),
\end{equation}%
with the decay rates $\gamma_a=2\pi(g^{a}_k)^2\rho(\omega_{ag})$,
$\gamma_b=2\pi(g_k^{b})^2\rho(\omega_{bg})$,
$\gamma_c=2\pi(g_k^{c})^2\rho(\omega_{cg})$,
and $\gamma_d=2\pi(g_k^{d})^2\rho(\omega_{dg})$, and mode density $\rho(\omega_k)$.
The dynamical equations (\ref{Eq3})-(\ref{Eq6}) can be rewritten with constant coefficients as
\begin{equation}\label{Eq11}
\frac{d}{dt}A\left( t\right) =-\frac{\gamma _{a}}{2}A\left( t\right)
-i\Omega _{ca}e^{i\Phi }\widetilde{C}\left( t\right) -i\Omega _{da}%
\widetilde{D}\left( t\right),
\end{equation}%
\begin{equation}\label{Eq12}
\frac{d}{dt}B\left( t\right) =-\frac{\gamma _{b}}{2}B\left( t\right)
-i\Omega _{cb}\widetilde{C}\left( t\right) -i\Omega _{db}\widetilde{D}\left(
t\right),
\end{equation}%
\begin{equation}\label{Eq13}
\frac{d}{dt}\widetilde{C}\left( t\right) =\left( -i\Delta _{c}-\frac{\gamma
_{c}}{2}\right) \widetilde{C}\left( t\right) -i\Omega _{ca}e^{-i\Phi
}A\left( t\right) -i\Omega _{cb}B\left( t\right),
\end{equation}%
\begin{equation}\label{Eq14}
\frac{d}{dt}\widetilde{D}\left( t\right) =\left( -i\Delta _{d}-\frac{\gamma
_{d}}{2}\right) \widetilde{D}\left( t\right) -i\Omega _{da}A\left( t\right)
-i\Omega _{db}B\left( t\right),
\end{equation}
with the definitions $\widetilde{C}\left( t\right) \equiv e^{-i\Delta _{c}t}C\left(
t\right) $, $\widetilde{D}\left( t\right) \equiv e^{-i\Delta _{d}t}D\left(
t\right) $.
It is worth mentioning that, here we have ignored the effects of the spontaneous decay between the four upper excited states, based on the assumption that the spontaneous relaxation between the four upper excited states is much slower than the spontaneous decay from the excited states to the ground states~\cite{HSongPRA19}, and also much slower than the coherent transition between the excited states induced by the strong driving fields.

\section{Transition probabilities} \label{TP}

\begin{figure}[tbp]
\centering
\includegraphics[bb=42 328 557 625, width=12 cm, clip]{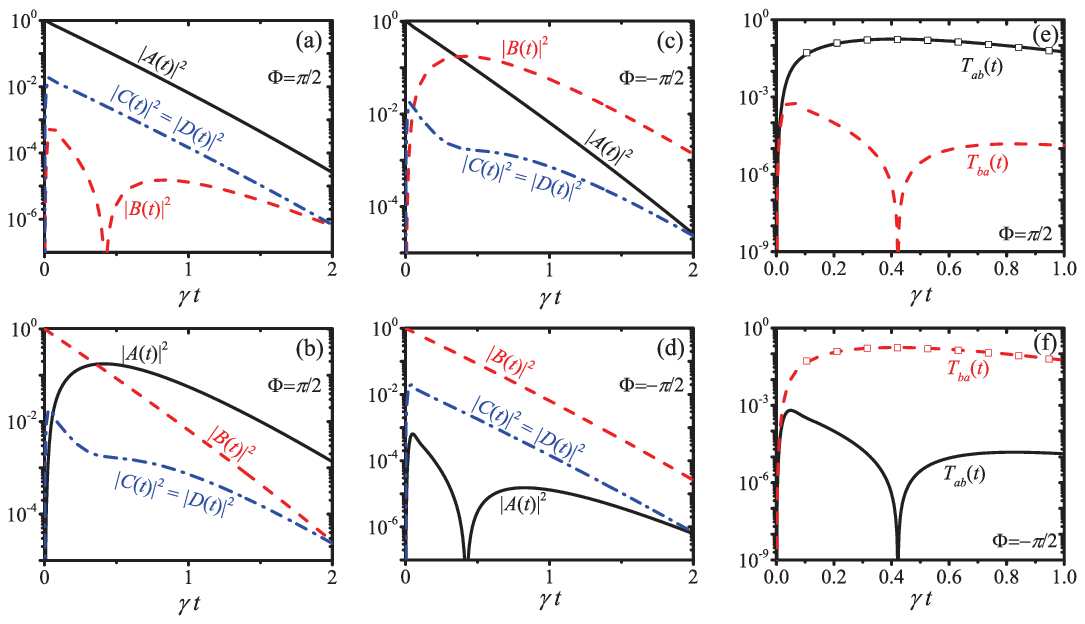}
\caption{(Color online) The populations $|A\left( t\right)|^2$ (black solid
curve), $|B\left( t\right)|^2$ (red dashed curve), and $|C\left( t\right)|^2=|D\left( t\right)|^2$ (blue dashed-dot curve) are plotted as
functions of the time $\gamma t$ for: $\Phi=\protect\pi/2$ in (a) and (b); $%
\protect\Phi=-\protect\pi/2$ in (c) and (d). The initial conditions are $A\left( 0\right)=1$ and $B\left( 0\right)=C\left( 0\right)=D\left( 0\right)=0$ for (a) and (c), $B\left( 0\right)=1$ and $A\left( 0\right)=C\left( 0\right)=D\left( 0\right)=0$
for (b) and (d). The transition probabilities $T_{ab}(t)$ (black solid
curve) and $T_{ba}(t)$ (red dashed curve) are plotted as functions of the time $\gamma t$ for: (e) $\Phi=%
\protect\pi/2$; (f) $\Phi=-\pi/2$. The other parameters are $\protect\gamma_a=\protect%
\gamma_b=\gamma$, $\protect\gamma_c=\gamma_d=100\gamma$, $\Omega_{ca}=\Omega_{cb}=
\Omega_{da}=\Omega_{db}=10\gamma$, $\Delta _{c}=-\Delta_{d}=50\gamma$.
}
\label{fig2}
\end{figure}

To gain insight into the transition probabilities between energy levels $|a\rangle$ and $|b\rangle$, we will solve the dynamic equations~(\ref{Eq11})-(\ref{Eq14}) both analytically and numerically.
The dynamic equations~(\ref{Eq11})-(\ref{Eq14}) can be rewritten as an effective Schr\"{o}dinger equation
\begin{equation}\label{Eq15}
i\frac{d\Psi \left( t\right) }{dt}=H_{\mathrm{eff}}\Psi \left( t\right) ,
\end{equation}%
with state vector $\Psi \left( t\right) =\left[ A\left( t\right) ,B\left( t\right) ,%
\widetilde{C}\left( t\right) ,\widetilde{D}\left( t\right) \right]
^{\intercal }$, and effective Hamiltonian as
\begin{equation}\label{Eq16}
H_{\mathrm{eff}}=\left(
\begin{array}{cccc}
-i\frac{\gamma _{a}}{2} & 0 & \Omega _{ca}e^{i\Phi } & \Omega _{da} \\
0 & -i\frac{\gamma _{b}}{2} & \Omega _{cb} & \Omega _{db} \\
\Omega _{ca}e^{-i\Phi } & \Omega _{cb} & \Delta _{c}-i\frac{\gamma _{c}}{2}
& 0 \\
\Omega _{da} & \Omega _{db} & 0 & \Delta _{d}-i\frac{\gamma _{d}}{2}%
\end{array}%
\right) .
\end{equation}%
The formal solutions for Eq.~(\ref{Eq15}) can be written as
\begin{equation}\label{Eq17}
\Psi \left( t\right) =U\left( t\right) \Psi \left( 0\right) ,
\end{equation}%
with initial vector $\Psi \left( 0\right) =\left[ A\left( 0\right)
,B\left( 0\right) ,C\left( 0\right) ,D\left( 0\right) \right] ^{\intercal }$%
, and time-evolution matrix%
\begin{equation}\label{Eq18}
U\left( t\right) \equiv e^{-iH_{\mathrm{eff}}t}.
\end{equation}%
The transition probabilities $T_{ba}(t)$ for $|a\rangle \rightarrow |b\rangle $
and $T_{ab}(t)$ for $|b\rangle \rightarrow |a\rangle $ can be defined from the time-evolution matrix $U\left( t\right)$ as
\begin{equation}\label{Eq19}
T_{ba}\left( t\right) \equiv \left\vert U_{21}\left( t\right) \right\vert
^{2},
\end{equation}%
\begin{equation}\label{Eq20}
T_{ab}\left( t\right) \equiv \left\vert U_{12}\left( t\right) \right\vert
^{2},
\end{equation}%
where the subscripts $1$ and $2$ are the row and column numbers of the matrix $U\left( t\right)$.
Moreover, the isolation for the nonreciprocal transition between levels $|a\rangle $ and $|b\rangle $ is defined by
\begin{equation}
  I(t)\equiv \frac{T_{ab}(t)}{T_{ba}(t)}.
\end{equation}

Under the assumption that the decay rates of the upper levels $|c\rangle$ and $|d\rangle$ are much larger than the other parameters, i.e., $\min \left\{ \gamma _{c},\gamma _{d}\right\} \gg \max
\left\{ \gamma _{a},\gamma _{b},\Omega _{ca},\Omega _{cb},\Omega
_{da},\Omega _{db}\right\} $, we can eliminate the auxiliary levels $|c\rangle$ and $|d\rangle$ adiabatically. By setting $d\widetilde{C}\left( t\right)/dt = d\widetilde{D}\left( t\right)/dt=0$ in Eqs.~(\ref{Eq13}) and (\ref{Eq14}), we obtain
\begin{eqnarray}
\widetilde{C}\left( t\right) &=&-\frac{i\Omega _{ca}}{ i\Delta _{c}+%
\frac{\gamma _{c}}{2} }e^{-i\Phi }A\left( t\right) -\frac{i\Omega
_{cb}}{ i\Delta _{c}+\frac{\gamma _{c}}{2} }B\left( t\right), \\
\widetilde{D}\left( t\right) &=&\frac{-i\Omega _{da}}{ i\Delta _{d}+%
\frac{\gamma _{d}}{2} }A\left( t\right) +\frac{-i\Omega _{db}}{
i\Delta _{d}+\frac{\gamma _{d}}{2} }B\left( t\right).
\end{eqnarray}%
Substitute these into Eqs.~(\ref{Eq11}) and (\ref{Eq12}), the effective dynamic equations for levels $|a\rangle $ and $|b\rangle $ can be written as
\begin{eqnarray}
\frac{d}{dt}A\left( t\right)  &=&-\left( \frac{\gamma _{a,\mathrm{eff}}}{2}%
+i\Delta _{a,\mathrm{eff}}\right) A\left( t\right) -J_{ab}B\left( t\right) ,
\\
\frac{d}{dt}B\left( t\right)  &=&-\left( \frac{\gamma _{b,\mathrm{eff}}}{2}%
+i\Delta _{b,\mathrm{eff}}\right) B\left( t\right) -J_{ba}A\left( t\right)
\end{eqnarray}%
with effective decay rates
\begin{equation}
\gamma _{a,\mathrm{eff}}=\gamma _{a}+\frac{4\gamma _{c}\Omega _{ca}^{2}}{%
4\Delta _{c}^{2}+\gamma _{c}^{2}}+\frac{4\gamma _{d}\Omega _{da}^{2}}{%
4\Delta _{d}^{2}+\gamma _{d}^{2}},
\end{equation}%
\begin{equation}
\gamma _{b,\mathrm{eff}}=\gamma _{b}+\frac{4\gamma _{c}\Omega _{cb}^{2}}{%
4\Delta _{c}^{2}+\gamma _{c}^{2}}+\frac{4\gamma _{d}\Omega _{db}^{2}}{%
4\Delta _{d}^{2}+\gamma _{d}^{2}},
\end{equation}%
effective detunings
\begin{equation}
\Delta _{a,\mathrm{eff}}=-\frac{4\Delta _{c}\Omega _{ca}^{2}}{4\Delta
_{c}^{2}+\gamma _{c}^{2}}-\frac{4\Delta _{d}\Omega _{da}^{2}}{4\Delta
_{d}^{2}+\gamma _{d}^{2}},
\end{equation}%
\begin{equation}
\Delta _{b,\mathrm{eff}}=-\frac{4\Delta _{c}\Omega _{cb}^{2}}{4\Delta
_{c}^{2}+\gamma _{c}^{2}}-\frac{4\Delta _{d}\Omega _{db}^{2}}{4\Delta
_{d}^{2}+\gamma _{d}^{2}},
\end{equation}%
and effective coupling coefficients
\begin{equation}
J_{ab}= \frac{\Omega _{ca}\Omega _{cb}e^{i\Phi }}{ i\Delta _{c}+%
\frac{\gamma _{c}}{2} }+\frac{\Omega _{da}\Omega _{db}}{
i\Delta _{d}+\frac{\gamma _{d}}{2} } ,
\end{equation}
\begin{equation}
J_{ba}= \frac{\Omega _{ca}\Omega _{cb}e^{-i\Phi }}{ i\Delta _{c}+%
\frac{\gamma _{c}}{2}}+\frac{\Omega _{da}\Omega _{db}}{
i\Delta _{d}+\frac{\gamma _{d}}{2} }.
\end{equation}
The condition for nonreciprocal transition is $J_{ab}\neq J_{ba}$, i.e., $\Phi \neq n\pi$ ($n$ is an integer).
This can be understood intuitively that $\Phi \neq n\pi$ breaks the time-reversal symmetry of the Hamiltonian given in Eq.~(\ref{Eq1}).

\begin{figure}[tbp]
\centering
\includegraphics[bb=150 282 429 611, width=8.5 cm, clip]{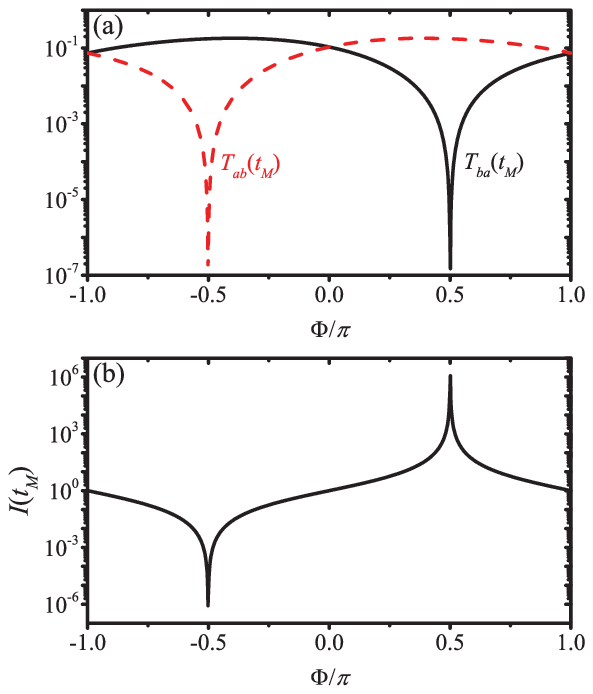}
\caption{(Color online) (a) The transition probabilities $T_{ba}(t)$ and $%
T_{ab}(t)$, and (b) the isolation $I(t)$ are plotted as functions of the
synthetic magnetic flux $\Phi$ at time $t=t_M$. The other parameters
are the same as in Fig.~\ref{fig2}.}
\label{fig3}
\end{figure}

One necessary condition for optimal nonreciprocal transition is one of the effective coupling coefficients ($J_{ab}$ or $ J_{ba}$) equal zero, i.e.,
\begin{equation}
e^{\pm i\Phi }=-\frac{\left( i2\Delta _{c}+\gamma _{c}\right) }{\left(
i2\Delta _{d}+\gamma _{d}\right) }\frac{\Omega _{da}\Omega _{db}}{\Omega
_{ca}\Omega _{cb}}.
\end{equation}
The transition probabilities from $|a\rangle $ to $|b\rangle $ [$T_{ba}(t)$]
and from $|b\rangle $ to $|a\rangle $ [$T_{ab}(t)$] can be obtained
approximately under this condition. For simplicity, we choose the symmetric
parameters: $\Omega _{ca}=\Omega _{cb}=\Omega _{da}=\Omega _{db}$, $\gamma
_{a}=\gamma _{b}$, $\gamma _{c}=\gamma _{d}$, $\Delta _{c}=-\Delta _{d}$.
The condition $J_{ab}=0$ ($J_{ba}=0$) is satisfied for $\Phi =-\pi /2$ ($\Phi =\pi /2$) with detuning $\Delta
_{c}=-\Delta _{d}=\gamma _{c}/2=\gamma _{d}/2$, and we have transition
probabilities%
\begin{equation}\label{Eq33}
T_{ba}(t)\approx \left\vert J_{ba}t\exp \left[ -\left( \frac{\gamma _{b,%
\mathrm{eff}}}{2}+i\Delta _{b,\mathrm{eff}}\right) t\right] \right\vert ^{2},
\end{equation}%
for initial conditions $A\left( 0\right)=1$ and $B\left( 0\right)=C\left( 0\right)=D\left( 0\right)=0$,
and
\begin{equation}\label{Eq34}
T_{ab}(t)\approx \left\vert J_{ab}t\exp \left[ -\left( \frac{\gamma _{a,%
\mathrm{eff}}}{2}+i\Delta _{a,\mathrm{eff}}\right) t\right] \right\vert ^{2}
\end{equation}%
for initial conditions $B\left( 0\right)=1$ and $A\left( 0\right)=C\left( 0\right)=D\left( 0\right)=0$.
The transition probabilities are time dependent, and the time for maximal
transition probability is
\begin{equation}
t_{M}\approx\frac{2}{\gamma _{b,\mathrm{eff}}}=\frac{2}{\gamma _{a,\mathrm{eff}}},
\end{equation}
with transition probabilities
\begin{equation}
T_{ba}(t_{M})\approx \left\vert \frac{2}{e}\frac{J_{ba}}{\gamma _{b,\mathrm{%
eff}}}\right\vert ^{2},
\end{equation}
\begin{equation}
T_{ab}(t_{M})\approx \left\vert \frac{2}{e}\frac{J_{ab}}{\gamma _{a,\mathrm{%
eff}}}\right\vert ^{2},
\end{equation}
where $e$ is the mathematical constant approximately equal to $2.71828$.

The populations $|A\left( t\right)|^2$ (black solid curve), $|B\left( t\right)|^2$ (red
dashed curve), and $|C\left( t\right)|^2=|D\left( t\right)|^2$ (blue dashed-dot curve) obtained from Eqs.~(\ref{Eq16})-(\ref{Eq18}) are plotted as
functions of the time $t$ in Figs.~\ref{fig2}(a)-\ref{fig2}(d).
It is clear that the population can transfer
from the level $|b\rangle$ to level $|a\rangle$ for $\Phi=\pi/2$, but
almost no population will transfer from the level $|a\rangle$ to level $%
|b\rangle$.
In contrast, the population can transfer
from the level $|a\rangle$ to level $|b\rangle$, but
almost no population will transfer from the level $|b\rangle$ to level $%
|a\rangle$ when $\Phi=-\pi/2$.

The transition probabilities from $|b\rangle$ to $|a\rangle$ [$T_{ab}(t)$] and from $|a\rangle$ to $|b\rangle$ [$T_{ba}(t)$] can also be obtained from Eqs.~(\ref{Eq18})-(\ref{Eq20}).
They are plotted as functions of time $ t$ in Figs.~%
\ref{fig2}(e) and \ref{fig2}(f). It is clear that $T_{ab}(t)\gg T_{ba}(t)$ for $%
\phi =\pi /2 $, and $T_{ab}(t)\ll T_{ba}(t)$ for $\phi =-\pi /2$,
i.e., the transitions between levels $|b\rangle$ and $|a\rangle$ are nonreciprocal.
The approximate analytical results given in Eqs.~(\ref{Eq33}) and (\ref{Eq34}) are shown by open squares in Figs.~%
\ref{fig2}(e) and \ref{fig2}(f), which agree well with the solid and dashed curves obtained from Eqs.~(\ref{Eq18})-(\ref{Eq20}).

Furthermore, the dependence of the transition probabilities $T_{ba}(t)$ and $%
T_{ab}(t)$ on the synthetic magnetic flux $\Phi$ is show in Fig.~\ref{fig3}%
(a). At time $t=t_M$, we have $T_{ba}(t)<T_{ab}(t)$ for synthetic
magnetic flux $0<\Phi<\pi$; in the contrast, we have $T_{ba}(t)> T_{ab}(t)$
for synthetic magnetic flux $-\pi<\Phi<0$. As shown in Fig.~\ref{fig3}(b),
under the conditions $\Omega _{ca}=\Omega _{cb}=\Omega _{da}=\Omega _{db}$ and $\Delta _{c}=-\Delta_{d}=\gamma_c/2=\gamma_d/2$, the
optimal isolation $I(t)$ is obtained with synthetic magnetic flux $\Phi =\pm
\pi /2$.

\section{Spontaneous emission spectrum} \label{SE}

\begin{figure}[tbp]
\centering
\includegraphics[bb=134 352 421 553, width=13 cm, clip]{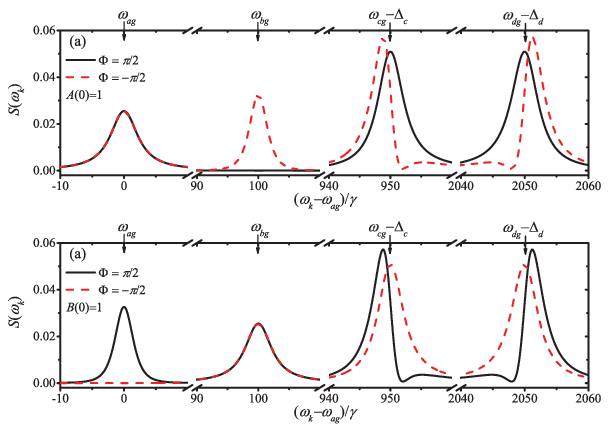}
\caption{(Color online) The spontaneous emission spectrum $S(\omega_k)$ is plotted as a function of the detuning $(\omega_k-\omega_{ag})/\gamma$ for synthetic magnetic flux $\Phi=\pi/2$ (black solid curve) and $\Phi=-\pi/2$ (red dashed curve), with initial conditions $A\left( 0\right)=1$ and $B\left( 0\right)=C\left( 0\right)=D\left( 0\right)=0$ in (a), and  $B\left( 0\right)=1$ and $A\left( 0\right)=C\left( 0\right)=D\left( 0\right)=0$ in (b). In both (a) and (b), we set $\omega_{bg}-\omega_{ag}=100\gamma$, $\omega_{cg}-\omega_{ag}=1000\gamma$, and $\omega_{dg}-\omega_{ag}=2000\gamma$. The other parameters are the same as in Fig.~\ref{fig2}.}
\label{fig4}
\end{figure}

In this section, we will show that the nonreciprocal transitions can be identified by measuring the spontaneous
emission spectra of the system.
The spontaneous emission spectra of the system can be derived analytically by the Laplace transform method~\cite{SYZhuPRA95,SYZhuPRL96}. Taking the Laplace transform, i.e., $\overline{O}%
\left( s\right) =\int_{0}^{+\infty }O\left( t\right) e^{-st}dt$,
the solution of the effective Schr\"{o}dinger equation~(\ref{Eq15}),
$\overline{\Psi }\left( s\right) =\left[ \overline{A}\left( s\right) ,%
\overline{B}\left( s\right) ,\overline{\widetilde{C}}\left( s\right) ,%
\overline{\widetilde{D}}\left( s\right) \right] ^{\intercal }$,
is given by
\begin{equation}\label{Eq38}
\overline{\Psi }\left( s\right) =M^{-1}\Psi\left( 0\right)
\end{equation}%
with coefficient matrix
\begin{equation}\label{Eq39}
M=\left(
\begin{array}{cccc}
s+\frac{\gamma _{a}}{2}  & 0 & i\Omega _{ca}e^{i\Phi } &
i\Omega _{da} \\
0 &  s+\frac{\gamma _{b}}{2}  & i\Omega _{cb} & i\Omega _{db}
\\
i\Omega _{ca}e^{-i\Phi } & i\Omega _{cb} &  s+\frac{\gamma _{c}}{2}%
+i\Delta _{c}  & 0 \\
i\Omega _{da} & i\Omega _{db} & 0 &  s+\frac{\gamma _{d}}{2}+i\Delta
_{d}
\end{array}%
\right) .
\end{equation}

The spontaneous emission spectrum of the system $S\left( \omega _{k}\right) =\sum_{i=a,b,c,d}\left\vert
G_{k}^{i}\left( +\infty \right) \right\vert ^{2}\rho \left(
\omega _{k}\right) $ is obtained from the Fourier transform of $\left\langle E^{-}\left( t+\tau \right) E^{+}\left( t\right) \right\rangle
_{t\rightarrow +\infty }$~\cite{SYZhuPRA95,SYZhuPRL96,XWXuArx20}, with $G_{k}^{i}\left( +\infty \right) \equiv \left.
G_{k}^{i}\left( t\right) \right\vert _{t\rightarrow +\infty }$ as the long
time behavior ($t\rightarrow +\infty $) of $G_{k}^{i}\left( t \right)$.
By integrating time $t$ in Eqs.~(\ref{Eq7})-(\ref{Eq10}), $G_{k}^{i}\left( +\infty \right)$ is given by
\begin{equation}
G_{k}^{a}\left( +\infty \right) =-ig_{k}^{a}\overline{A}\left[ i\left( \omega _{ag}-\omega
_{k}\right) \right] ,
\end{equation}%
\begin{equation}
G_{k}^{b}\left( +\infty \right) =-ig_{k}^{b}\overline{B}\left[ i\left( \omega _{bg}-\omega
_{k}\right) \right] ,
\end{equation}%
\begin{equation}
G_{k}^{c}\left( +\infty \right) =-ig_{k}^{c}\overline{\widetilde{C}}\left[ i\left( \omega
_{cg}-\Delta _{c}-\omega _{k}\right) \right] ,
\end{equation}%
\begin{equation}
G_{k}^{d}\left( +\infty \right) =-ig_{k}^{d}\overline{\widetilde{D}}\left[ i\left( \omega
_{dg}-\Delta _{d}-\omega _{k}\right) \right] .
\end{equation}%
Thus, we find the spontaneous emission spectrum as
\begin{eqnarray}
S\left( \omega _{k}\right)  &=&\frac{\gamma _{a}}{2\pi }\left\vert \overline{%
A}\left[ i\left( \omega _{ag}-\omega _{k}\right) \right] \right\vert ^{2}+%
\frac{\gamma _{b}}{2\pi }\left\vert \overline{B}\left[ i\left( \omega
_{bg}-\omega _{k}\right) \right] \right\vert ^{2}  \nonumber \\
&&+\frac{\gamma _{c}}{2\pi }\left\vert \overline{\widetilde{C}}\left[
i\left( \omega _{cg}-\Delta _{c}-\omega _{k}\right) \right] \right\vert ^{2}
 \nonumber \\
&&+\frac{\gamma _{d}}{2\pi }\left\vert \overline{\widetilde{D}}\left[ i\left(
\omega _{dg}-\Delta _{d}-\omega _{k}\right) \right] \right\vert ^{2}
\end{eqnarray}%
with $\overline{A}\left( s\right) $, $\overline{B}\left( s\right) $, $%
\overline{\widetilde{C}}\left( s\right) $, and $\overline{\widetilde{D}}%
\left( s\right) $ given by Eqs.~(\ref{Eq38}) and (\ref{Eq39}).

\begin{figure}[tbp]
\centering
\includegraphics[bb=14 4 301 169, width=8.5 cm, clip]{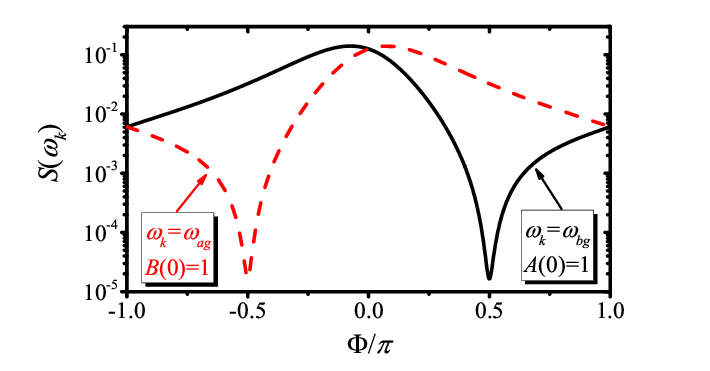}
\caption{(Color online) The spontaneous emission spectrum $S(\omega_k)$ is plotted as a function of the
synthetic magnetic flux $\Phi$: (black solid curve) with frequency $\omega_k=\omega_{bg}$ and initial conditions $A\left( 0\right)=1$ and $B\left( 0\right)=C\left( 0\right)=D\left( 0\right)=0$, and (red dashed curve) with frequency $\omega_k=\omega_{ag}$ and initial conditions $B\left( 0\right)=1$ and $A\left( 0\right)=C\left( 0\right)=D\left( 0\right)=0$. The other parameters
are the same as in Fig.~\ref{fig4}.}
\label{fig5}
\end{figure}

In Fig.~\ref{fig4}, the spontaneous emission spectra of the system are plotted with different initial conditions: (a) $A\left( 0\right)=1$ and $B\left( 0\right)=C\left( 0\right)=D\left( 0\right)=0$; (b) $B\left( 0\right)=1$ and $A\left( 0\right)=C\left( 0\right)=D\left( 0\right)=0$.
In Fig.~\ref{fig4} (a), we assume that the system is prepared in level $|a\rangle$ initially, and the black solid curves are for phase $\Phi=\pi/2$ and the red dashed curves for $\Phi=-\pi/2$.
When $\Phi=-\pi/2$, as the population can
transfer from the level $|a\rangle$ to level $|b\rangle$, so there is a
peak around the resonant frequency $\omega_{bg}$ in the spontaneous emission spectra.
In the meantime, the population can
transfer from both the levels $|a\rangle$ and $|b\rangle$ to levels $|c\rangle$ and $|d\rangle$, and the interferences between different population transferring paths suppress the population decay from levels $|c\rangle$ and $|d\rangle$ [see $|C\left( t\right)|^2=|D\left( t\right)|^2$ (blue dashed-dot curve) in Fig.~\ref{fig2}(c)] and induce the dips around frequencies $\omega_{cg}-\Delta_c$ and $\omega_{bg}-\Delta_d$.
Instead, there is almost no population transferring from the level $|a\rangle$ to level $|b\rangle$ when $%
\Phi=\pi/2$, so that the peak around the frequency $\omega_{bg}$ is eliminated.
Meanwhile, the population transfer only from the level $|a\rangle$ to levels $|c\rangle$ and $|d\rangle$, so that the population in levels $|c\rangle$ and $|d\rangle$ decay nearly exponentially [see $|C\left( t\right)|^2=|D\left( t\right)|^2$ (blue dashed-dot curve) in Fig.~\ref{fig2}(a)] and there are single peaks around the frequencies $\omega_{cg}-\Delta_c$ and $\omega_{bg}-\Delta_d$.
In Fig.~\ref{fig4} (b), we assume that the system is prepared in level $|b\rangle$ initially.
When $\Phi=\pi/2$, as the population can
transfer from the level $|b\rangle$ to level $|a\rangle$, so there is a
peak around the resonant frequency $\omega_{ag}$
and dips around frequencies $\omega_{cg}-\Delta_c$ and $\omega_{bg}-\Delta_d$
in the spontaneous emission spectra. Instead, there is almost no
population transferring from the level $|b\rangle$ to level $|a\rangle$ when $%
\Phi=-\pi/2$, so that the peak around the frequency $\omega_{ag}$ is
eliminated and there are single peaks around the frequencies $\omega_{cg}-\Delta_c$ and $\omega_{bg}-\Delta_d$.

In addition, the dependence of the
spontaneous emission spectra $S(\omega_k=\omega_{ag})$ and $S(\omega_k=\omega_{bg})$ on the synthetic magnetic flux $\Phi$ are shown in Fig.~\ref{fig5}, with the system initially prepared in level $|b\rangle$ and $|a\rangle$ respectively. We have $S(\omega_k=\omega_{ag})>S(\omega_k=\omega_{bg})$ for synthetic
magnetic flux $0<\Phi<\pi$, which corresponds with $T_{ba}(t)<T_{ab}(t)$ in Fig.~\ref{fig3}(a); in the contrast, we have $S(\omega_k=\omega_{ag})<S(\omega_k=\omega_{bg})$ for synthetic magnetic flux $-\pi<\Phi<0$, conforming to $T_{ba}(t)> T_{ab}(t)$. Thus we can also observe the nonreciprocal transitions by contrasting the difference of the spontaneous emission spectra $S(\omega_k=\omega_{ag})$ and $S(\omega_k=\omega_{bg})$ with the system prepared in level $|b\rangle$ or $|a\rangle$ initially.
Moreover, almost all the populations have decayed to the ground states at the time $t=2/\gamma$, as shown in Figs.~\ref{fig2}(a)-\ref{fig2}(d), so the elimination of spectral lines can be observed within the time scale of $2/\gamma$.

\section{Conclusions} \label{CN}

Let us now discuss the experimental feasibility of our proposal.
Take $^{40}{\rm Ca}^+$ for example, let us show the corresponding relationship between the energy levels in Fig.~\ref{fig1} and the energy levels of $^{40}{\rm Ca}^+$: $|a\rangle=|3D_{5/2}\rangle$, $|b\rangle=|3D_{3/2}\rangle$, $|c\rangle=|5P_{3/2}\rangle$, $|d\rangle=|4P_{3/2}\rangle$.
First of all, the one-photon dipole transitions $|4P_{3/2}\rangle\leftrightarrow |3D_{5/2}\rangle$, $|4P_{3/2}\rangle\leftrightarrow |3D_{3/2}\rangle$, $|5P_{3/2}\rangle\leftrightarrow |3D_{5/2}\rangle$, and $|5P_{3/2}\rangle\leftrightarrow |3D_{3/2}\rangle$ are allowed, but the one-photon dipole transitions $|5P_{3/2}\rangle\leftrightarrow |4P_{3/2}\rangle$ and $ |3D_{5/2}\rangle\leftrightarrow |3D_{3/2}\rangle$ are forbidden, i.e., the cyclic transitions for the four levels can be realized.
To avoid the effects of Doppler broadening, the calcium ions should be laser cooled primarily.
A single calcium ion can be trapped and laser cooled to 1 mK by Doppler cooling in experiments~\cite{BartonPRA00}.
According to the experiments with single cold calcium ions~\cite{BartonPRA00,KreuterPRA05,HShaoPRA16,MeirPRA20} or few cold calcium ions~\cite{StaanumPRA04}, $|3D_{5/2}\rangle$ and $ |3D_{3/2}\rangle$ are metastable levels with lifetimes about $1.2$ seconds~\cite{BartonPRA00,StaanumPRA04,KreuterPRA05,HShaoPRA16}, the lifetimes of $|4P_{3/2}\rangle$ is about $6.6$ nanoseconds~\cite{JJinPRL93,MeirPRA20}, and the lifetimes of $|5P_{3/2}\rangle$ must also be much shorter than $1.2$ seconds, i.e., the conditions for adiabatic approximation ($\min \left\{ \gamma _{c},\gamma _{d}\right\} \gg \max
\left\{ \gamma _{a},\gamma _{b}\right\}$) are satisfied.
Hence if four laser fields with appropriate frequencies, strengths, and phases are applied to the $^{40}{\rm Ca}^+$ system, the proposal is realizable.
Moreover, similar energy structures can be found in other singly charged alkaline-earth metals ions, such as $^{88}{\rm Sr}^+$~\cite{MannervikPRL99,LetchumananPRA05} and $^{138}{\rm Ba}^+$~\cite{NYuPRL97,DijckPRA18,GurellPRA07,HaveyPRA77}.

In summary, a theoretically scheme was proposed to realize nonreciprocal transition between two indirectly coupled energy levels in a multi-level atomic system with cyclic four-level configuration. The spontaneous emission spectra of the multi-level system with nonreciprocal transition was also investigated. The nonreciprocal transition results in the elimination of a spectral line in the spontaneous emission spectrum, which can be used to identify the nonreciprocal transition experimentally.
As the multi-level atomic system with cyclic four-level configuration can be realized under the rules of electric dipole transitions, the scheme in this paper can be applied for observing nonreciprocal transition in the natural atoms or ions with parity symmetry, which will widely broaden the application sphere of nonreciprocal transition.

\vskip 2pc \leftline{\bf Acknowledgement}

X.-W.X. and H.-Q.S. are supported by the National Natural Science Foundation of China (NSFC) under Grant No.~12064010, Natural Science Foundation of Hunan Province of China under Grant No.~2021JJ20036, and the Natural Science Foundation of Jiangxi Province of China under Grant No.~20192ACB21002.
A.-X.C. is supported by NSFC under Grant No.~11775190.


\bibliography{ref}

\end{document}